# Topological Insulator Metamaterial with Giant Circular Photogalvanic Effect

X. Sun[1,2,†], G. Adamo[1,2,†], M. Eginligil[2,‡], H. N. S. Krishnamoorthy[1,2], N. I. Zheludev[1,2,3] and C. Soci[1,2,*]

1. Centre for Disruptive Photonic Technologies, TPI, SPMS, Nanyang Technological University, Singapore 637371, Singapore
2. Division of Physics and Applied Physics, Nanyang Technological University, Singapore 637371, Singapore
3. Optoelectronics Research Centre & Centre for Photonic Metamaterials, University of Southampton, Southampton SO17 1BJ, UK

†. These authors contributed equally
‡. Current address: Key Laboratory of Flexible Electronics (KLOFE) & Institute of Advanced Materials (IAM), Nanjing Tech University, 30 South Puzhu Road, Nanjing 211816, China
*Correspondence: csoci@ntu.edu.sg

**One of the most striking manifestations of electronic properties of topological insulators is the dependence of the photocurrent direction on the helicity of circularly polarized optical excitation. The helicity dependent photocurrents, underpinned by spin-momentum locking of surface Dirac electrons, are weak and easily overshadowed by bulk contributions. Here we show that the chiral response can be enhanced by nanostructuring. The tight confinement of electromagnetic fields in the resonant nanostructure enhances the photoexcitation of spin-polarized surface states of topological insulator $Bi_{1.5}Sb_{0.5}Te_{1.8}Se_{1.2}$, leading to an 11-fold increase of the circular photogalvanic effect and an unprecedented photocurrent dichroism ($\rho_{circ}$=0.87) at room temperature. The control of spin-transport in topological materials by structural design is a previously unrecognised ability of metamaterials that bridges the gap between nanophotonics and spin-electronics, providing new opportunities for developing polarization sensitive photodetectors.**

Chirality, or the difference of an object from its mirror image, is a ubiquitous and fascinating phenomenon in nature. It manifests itself at a variety of scales and forms, from galaxies to nanotubes, from organic molecules, to inorganic compounds. Detection of chirality at the molecular or atomic level is key to fundamental sciences (e.g., chemistry, biology, crystallography) and practical applications (e.g., food and pharmaceutical industry), yet very challenging. Detection of chirality relies on the interaction with electromagnetic fields, which is hindered by the large mismatch between the wavelength of light and the size of most molecules and crystalline unit cells, thereby resulting in nearly imperceptible twists of the light field over nanoscale dimensions.

In recent years, the chiral response of topological materials, such as topological insulators[1-4], Dirac/Weyl semimetals[5-8], and 2D van der Waals heterostructures[9,10], underpinned by the peculiar quantum features of their electronic structures, have been the subject of intense investigation. For instance, the surfaces of three-dimensional topological insulators support chiral spin currents, a net directional flow of spins in the absence of a net flow of charges. While the spin-texture of surface electrons in topological insulators can be unravelled by scanning tunnelling spectroscopy (STM)[11,12] or angular resolved photoemission spectroscopy (ARPES)[1,13,14], direct electrical or optical addressing of chiral surface states in this class of materials is hampered by the large contribution of the semiconducting bulk to the conductivity, induced by unintentional doping[15-18].

One of the most striking manifestations of the electronic chirality of topological insulator surface states is the dependence of their photocurrent on light helicity. Spin currents can be transformed into polarized net electrical currents when the system is optically driven out-of-equilibrium[2,19] by the helicity of light excitation: the absorption of circularly polarized light incident on the surface at oblique angles induces an unbalance between surface carriers of opposite spin at specific points in the *k*-space, thereby generating topological, spin-momentum locked photocurrents.

Attempts to magnify helicity-dependent photocurrent (HDPC) in topological materials have relied on strategies to increase the inherent surface-to-bulk contribution, such as using Bi-chalcogenides with low intrinsic doping[3,20-22], reducing the crystal thickness to just a few quintuple layers[23], tuning the Fermi energy by electrical gating[24], or selectively exciting surface carriers below the bulk band gap[25]. More recently, research has broadened to Weyl semimetals, where the "chiral anomaly"[6,7,26], induced by unpaired Dirac cones, is expected to yield highly chiral photocurrents.

Designer metamaterials, with structural features comparable to the wavelength of light, provide an independent approach to devise optical properties on demand and enhance light-matter interaction[27]. The creation and enhancement of optical chirality by metamaterials is particularly interesting and elaborate[28-31]. For instance, patterning an achiral medium (e.g. a metal) with chiral, subwavelength shapes makes the resulting metamaterial *intrinsically chiral*. Similarly, the mutual orientation of a metamaterial, patterned with non-chiral structures, and the incident electromagnetic wave can lead to an *extrinsically chiral* response[32,33]. Both forms of chirality can emerge in 2D and 3D metamaterial systems, and lead to strong optical activity, circular dichroism and asymmetric transmission.

Here we provide a first demonstration of the use of artificial nanostructuring to enhance the chiral photo-galvanic response of topological insulators (BSTS). Thanks to the tight confinement of electromagnetic fields, resonant non-chiral metamaterials effectively enhance the photoexcitation of spin-polarized states, thereby increasing the unbalance between surface-state carriers with opposite spin helicity, and overall resulting in a giant enhancement of the extrinsic chiral photocurrent response of a 3D topological insulator (TI).

Surface carriers can be selectively excited in topological insulators by circularly polarized light directed at oblique incidence on the surface of the crystal, and the resulting current flow is determined by the spin-momentum locking of the carriers (**Fig. 1**). As the topological insulator crystal is intrinsically achiral (the surface states have equal number of carriers with opposite spin orientations), and the relevant spins lay in the surface plane of the crystal, photoexcitation at normal incidence does not generate any HDPC. However, spin-selective photoexcitation of surface state carriers by obliquely incident light with a given helicity induces chirality (unbalance in the populations of surface state carriers with opposite spin orientations), akin the extrinsic chirality described for metamaterials. This effect, known as Circular Photogalvanic Effect (CPGE)[34-36], is illustrated in Fig. 1a, where a light beam of defined handedness selectively promotes carriers of matching spin from surface to bulk conduction bands, thus leaving the surface bands asymmetrically depopulated in *k*-space: the excess surface carriers of opposite spin generate the net spin-polarized electrical current, $j_a$. The presence of a nanostructure on the surface of topological insulators, designed to have a resonant absorption at the wavelength of excitation, effectively increases the number of surface conduction carriers which are promoted to the bulk conduction bands (Fig. 1b), thus significantly enhancing the CPGE contribution to the photocurrent. Fig. 1c shows the experimental arrangement to measure helicity-dependent photocurrent: light is incident at a variable angle, θ, on the surface of the topological insulator crystal (x-y plane), and the light polarization is changed continuously from linear to circular by rotating a quarter wave plate, whose fast axis forms an angle, φ, with the polarization axis. Without any applied bias, a helicity-dependent photocurrent flows across the two gold contacts on the crystal surface.

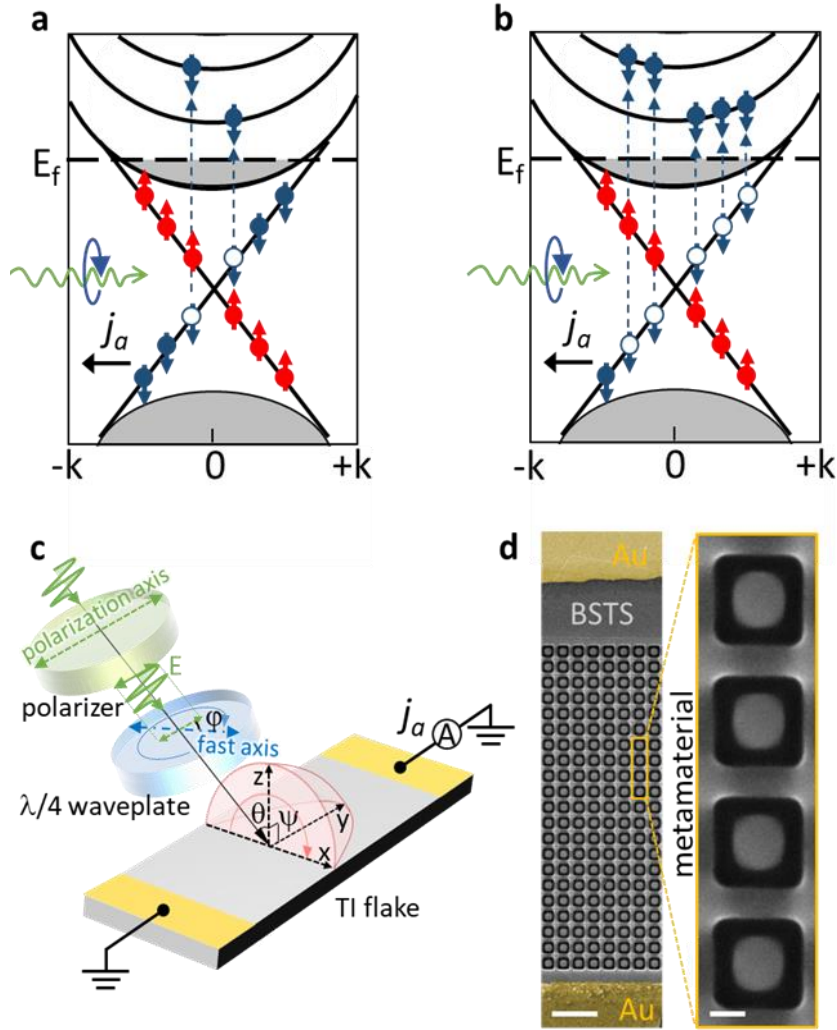

**Figure 1: Helicity dependent photocurrent in topological insulators and topological insulator metamaterials. a.** In an unstructured topological insulator (TI), Dirac electrons with spin coupled to a given circular polarization of incident light (blue) are promoted to higher bands in the *k*-space; the excess of spin-momentum locked surface-state electrons with opposite spin (red) gives rise to an helicity-dependent photocurrent, $j_a$ (circular photogalvanic effect). **b.** In a TI metamaterial, a larger number of spin-polarized electrons is photoexcited upon resonant light absorption, enhancing the helicity dependent photocurrent (HDPC). **c.** Schematic of the HDPC experimental setup, illustrating the mutual orientation of the electrodes on the TI device relative to the laser excitation beam at incidence angle θ and polarization defined by the angle of rotation φ of the quarter waveplate. **d.** Scanning electron microscope image of the square ring metamaterial carved between the Au electrodes on the surface of a TI flake (scale bars are 1 μm on the left and 100 nm on the right).

To assess the enhancement of chirality exclusive to topological surface spin currents, we chose a metamaterial design which does not introduce optical chirality, neither intrinsic or extrinsic[37]. The metamaterial unit cells consist of square rings of ~200 nm lateral size and ~100 nm ring width, arranged on a square lattice of ~335 nm period. Such pattern was carved by focused ion beam (FIB) milling between two gold electrodes deposited on a ~250 nm thin BSTS flake, as shown in the scanning electron microscope (SEM) images in Fig. 1d.

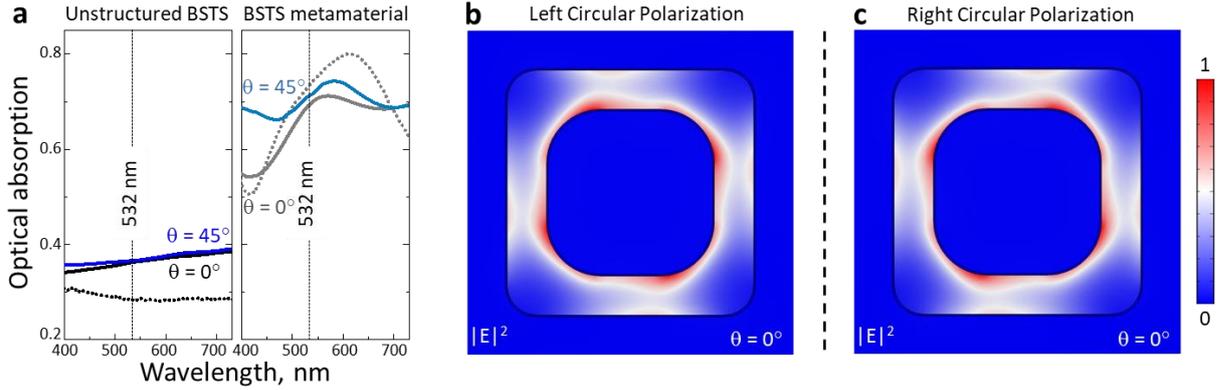

**Figure 2: Optical absorption enhancement in BSTS achiral metamaterial. a.** Measured (dashed lines) and simulated (continuous lines) optical absorption of an unstructured BSTS flake and a nanostructured BSTS metamaterial (experimental data were collected with unpolarized light at normal incidence and NA=0.7, while simulations correspond to circularly polarized light incident at θ=0° and θ=45°); at λ=532 nm, the absorption of the BSTS metamaterial is ~0.7, twice that of the unstructured BSTS flake (~0.35). **b-c.** Maps of the electric field intensity, $|E|^2$, at 10 nm below the top surface of the metamaterial unit cell at normal incidence, for left (LCP) and right (RCP) circular polarization, respectively.

The metamaterial was designed to resonantly increase the optical absorption of the unstructured BSTS flake at the excitation wavelength, λ=532 nm, at both normal ($\theta = 0°$) and oblique ($\theta = 45°$) incidence (**Fig. 2**). The experimental and numerically simulated spectra of the BSTS metamaterial (Fig. 2a, right panel) show a resonant increase in optical absorption with respect to the case of unstructured BSTS (Fig. 2a, left panel). Following from the design symmetry, the metamaterial geometry does not induce any helicity dependence to the bulk response of the BSTS flake, as confirmed by the maps of electric field intensity for circularly polarized light of opposite handedness (Figs. 2b and 2c).

We measured HDPC under nearly uniform illumination (laser beam diameter of ~200 μm, much larger than the BSTS device size of ~10 μm), with no applied bias (**Fig. 3**). The residual un-uniformity of illumination results in thermal gradient that induces polarization-independent photo-thermoelectric currents, which contribute to the photocurrent background. This component of the current is sensitive to the position of the excitation beam on the sample[3]. To measure a clearly distinguishable photocurrent signal, we consistently positioned the laser beam near the centre of the BSTS flake and maximized the total photocurrent response (see Supplementary Information, Section I, **Fig. S1**). Contributions of surface carriers to the photocurrent, seen in HDPC, are allowed only at oblique illumination[2], as illustrated in the top section of Fig. 3a. We measured the photocurrent, $j_a$, shining the laser at $\theta = -45°$ and going through a full rotation cycle of its polarization angle, $\varphi$, from 0° to 360°. This induces a continuous change of the incident polarization, from linear ($\varphi = 0°, 90°$) to circular of right ($\varphi = 45°$) and left ($\varphi = 135°$) handedness, with 180° period.

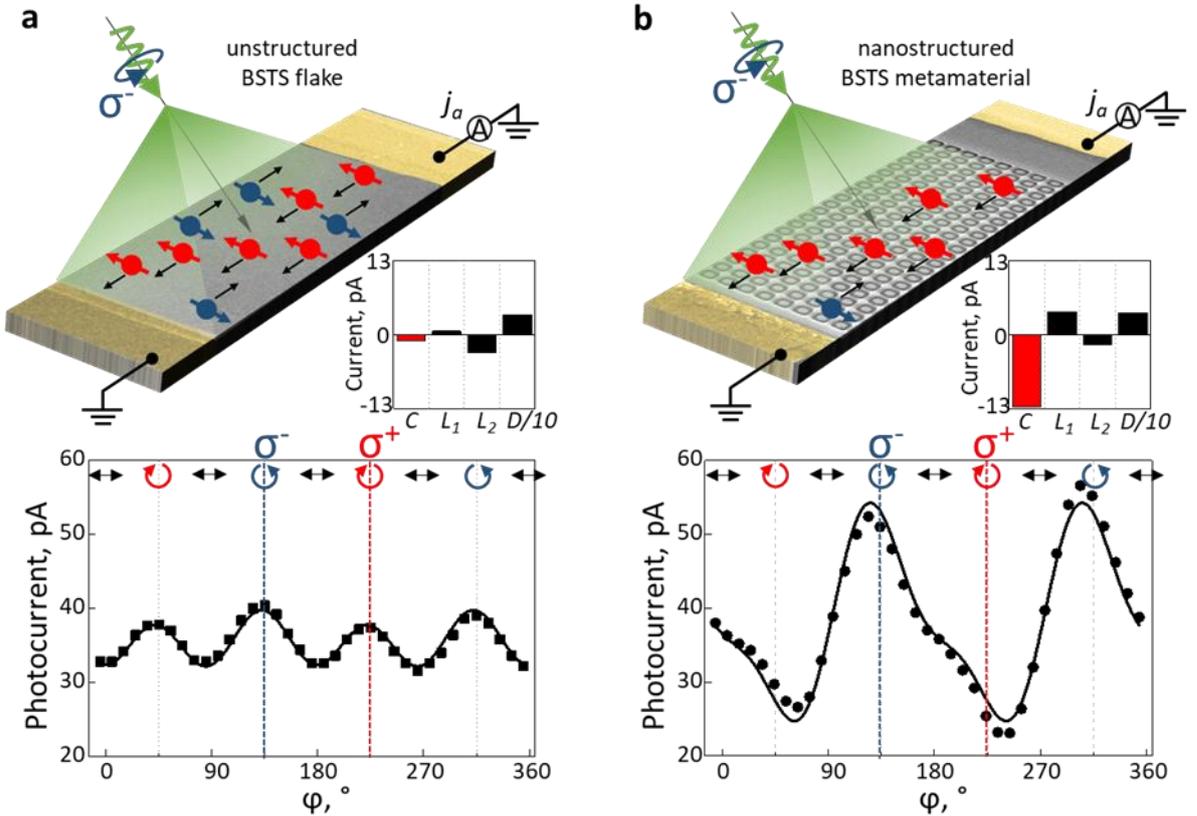

**Figure 3: Multifold increase of CPGE in BSTS topological insulator by metamaterials. a.** (top) Schematic of helicity dependent photocurrent (HDPC) in an unstructured BSTS flake; (bottom) experimental photocurrent measured on an unstructured BSTS flake, at room temperature, and fitting with Eq. (1), showing the expected 4φ dependence and a small 2φ asymmetry between right (σ⁺) and left (σ⁻) circularly polarized illumination; (inset) fitting coefficient $C$, $L_1$, $L_2$ and $D$, indicating a predominance of bulk photon-drag contribution, $L_2$, in the photocurrent modulation. **b.** (top) Schematic of light helicity dependent photocurrent (HDPC) in a nanostructured BSTS metamaterial; (bottom) experimental photocurrent measured on a nanostructured BSTS metamaterial, at room temperature, and fitting with Eq. (1), showing how the metamaterial induces a 2φ dependence which indicates that the sample responds almost exclusively to right (σ⁺) and left (σ⁻) circularly polarized illumination; (inset) fitting coefficient $C$, $L_1$, $L_2$ and $D$, indicating a predominance of CPGE, $C$, in the photocurrent modulation.

The bottom panel of Fig. 3a shows the photocurrent, $j_a$, measured in an unstructured BSTS flake (black filled squares). The current has the characteristic polarization dependent behaviour observed in other Bi-chalcogenide topological insulators[3,20,24] and other 2D material systems (i.e. quantum wells[34,35], transition metal dichalcogenide[8,38,39] and Weyl semimetals[6,7,26]). It comprises of four components, expressed by the phenomenological equation[3,35]:

$$j_a = C \sin 2\varphi + L_1 \sin 4\varphi + L_2 \cos 4\varphi + D \quad (1)$$

The coefficients D and $L_2$, are associated to photocurrent contributions from the semiconducting bulk. Specifically, D is related to the polarization independent photo-thermoelectric background current, which sets the overall directional current flow, while $L_2$ to the photon drag effect, which results from linear momentum transfer of the incident photons to the excited carriers. Conversely, the coefficients $L_1$ and C quantify photocurrent contributions from surface carriers. Such currents, driven by linear and circular polarization of the incident light, originate from the linear and circular photogalvanic effects, respectively. The dependence of the photocurrent on helicity is seen in the small asymmetry between the peaks corresponding to photoexcitation by light of opposite handedness ($\sigma^+$ and $\sigma^-$ in the bottom panel of Fig. 3a). The relative contribution of each surface and bulk component to the total photocurrent is shown in the bar plot of the coefficients of Eq. 1 (inset of Fig. 3a). While surface contributions to the photocurrent are discernible even at room temperature in the unstructured BSTS samples (thanks to the large surface to bulk conductivity known for this particular stoichiometry[22,40-42]), their dependence on light helicity is overshadowed by bulk components (*C/D*=0.03), and too small for any practical device or application.

The very same BSTS flake, patterned with square ring metamaterial array, behaves dramatically differently. The resonant metamaterial structure induces much larger asymmetry in the population of surface conducting bands, increasing the net spin current (top schematic in Fig. 3b). The effect is so strong that the measured photocurrent (black filled circles in bottom panel of Fig. 3b) has a distinct $2\varphi$ evolution as function of polarization, irrespective of the contribution of linearly polarized components[43]. The effect is even more apparent when comparing the coefficients of Eq. 1 (inset of Fig. 3b), where now *C* is far higher than the linear coefficients, $L_1$ and $L_2$, and its value accounts for a significant fraction of the photocurrent (*C/D*=0.33). **Table 1** summarizes the fitting coefficients of photocurrents, $j_a$, for both unstructured and nanostructured BSTS, at and off normal incidence. The photocurrent circular dichroism induced by spin-polarized surface states can be defined as[35]:

$$\rho_{circ} = \frac{|I_{\sigma^+} - I_{\sigma^-}|}{|I_{\sigma^+} + I_{\sigma^-}|} \quad (2)$$

where $I_{\sigma+}$ and $-I_{\sigma-}$ are respectively the values of photocurrents under left and right circularly polarized optical excitation, excluding the spin insensitive component $D$[44]. A 3-fold increase of the degree of spin polarization of unstructured BSTS ($\rho_{circ}$=0.26) is seen in the BSTS metamaterial ($\rho_{circ}$=0.87), an unprecedented degree of spin polarization approaching unity even at room temperature in non-magnetic materials[44-47]. Additional validation of the functional dependence of the HDPC on incidence angle θ, is shown in Supplementary Information (Section II, **Fig. S2** and **Fig. S3**). Notably, the surface nature of HDPC in BSTS was also confirmed repeating the experiments on a trivial chalcogenide insulator of the same family, $Bi_2S_3$, where we measured only bulk currents (Section III, **Fig. S4** and **Fig. S5**).

**Table 1: Fitting coefficients of the helicity-dependent photocurrent in unstructured BSTS flake and nanostructured BSTS metamaterials.**

| θ (°) | BSTS flake | $C$ (pA) | $L_1$ (pA) | $L_2$ (pA) | $D$ (pA) |
|---|---|---|---|---|---|
| -45 | unstructured | -1.1 ±0.1 | 0.6 ±0.1 | -3.1 ±0.1 | 35.4 ±0.1 |
|  | nanostructured | -12.7 ±0.3 | 4 ±0.3 | -1.8 ±0.3 | 38.3 ±0.3 |
| 0 | unstructured | 0.05 ±0.05 | -0.002 ±0.05 | 0.09 ±0.05 | 8.5 ±0.05 |
|  | nanostructured | -0.2 ±0.04 | -0.2 ±0.04 | 0.04 ±0.04 | 11.4 ±0.04 |
| 45 | unstructured | 1.0 ±0.1 | 0.32 ±0.1 | -3.2 ±0.1 | 35.1 ±0.1 |
|  | nanostructured | 11 ±0.4 | -0.9 ±0.4 | -3 ±0.4 | 41.8 ±0.4 |

For a given angle of incidence, the $j_a$ coefficients of both, unstructured and nanostructured BSTS, have equal sign, which reverses at mirror angles of incidence (Table 1 and Supplementary Information, Fig. S2). This proves that the metamaterial does not introduce chirality[37,48] but rather enhances the extrinsic chirality of the BSTS surface layer.

In the following, we report a first attempt to describe the photocurrent behaviour of spin-transport metamaterials by electromagnetic modelling. The generated photocurrent is directly proportional to the optical absorption, carrier density, mobility and lifetime of the topological insulator. Thus, under the assumption that the optical absorption of the BSTS metamaterial increases upon nanostructuring, while the remaining transport parameters remain unaltered, carrier anisotropy can be mapped to an anisotropic optical model of the BSTS topological insulator crystal (**Fig. 4**). Here we describe the unstructured BSTS crystal by its isotropic relative permittivity, $\varepsilon_r$ (experimentally determined by ellipsometric measurements[49]), modified by *ad-hoc* off-diagonal terms of the permittivity tensor to mimic the effective optical chirality induced by the in-plane spin of Dirac surface electrons. We performed full-wave electromagnetic simulations for both unstructured and nanostructured BSTS, replicating the sample illumination conditions used in the experiments ($\lambda = 532$ nm, $\theta = 0, \pm 45°$, $\varphi = 0°$ to $360°$), and integrated the electric field intensity within the top 3 nm[50] to evaluate the optical absorption at the surface of the topological insulator.

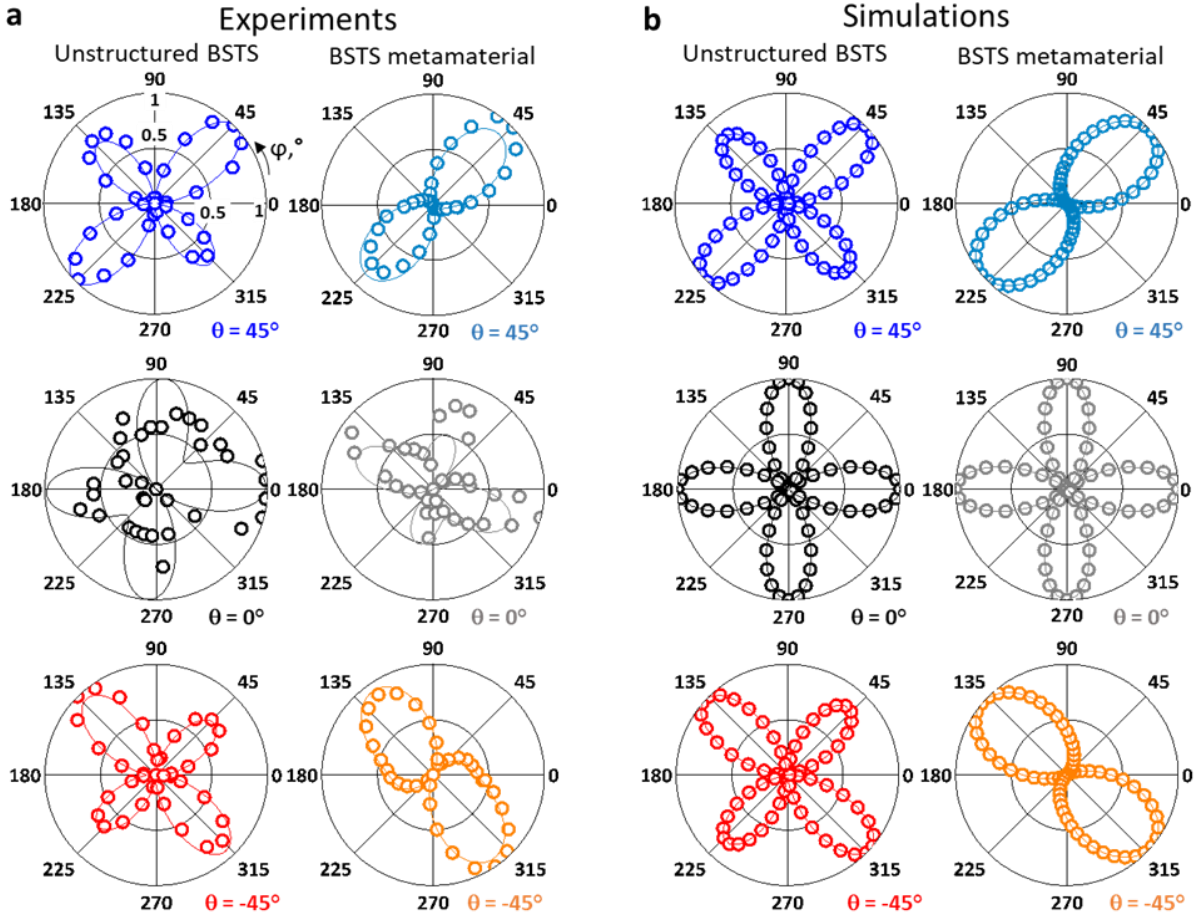

**Figure 4: Distillation of CPGE in BSTS topological insulator flakes by nanostructured metamaterials. a.** Normalized polar plots of helicity dependent photocurrent (HDPC) for an unstructured BSTS flake (left column) and a nanostructured BSTS metamaterial (right column) at 3 different angle of incidence, $\theta=45°$ (top row), $\theta=0°$ (centre row) and $\theta=-45°$ (bottom row); at $\theta=0°$ mostly $L_2$ contributes to the modulation of the photocurrent; at $\theta=45°$ and $\theta=-45$ the HDPC patterns of the unstructured BSTS flake and the BSTS metamaterial are distinctively different: the metamaterial distils the C term contribution to the photocurrent modulation, with respect to the unstructured BSTS flake, where $L_1$, $L_2$, and C have comparable amplitudes. **b.** Simulated $|E|^2$ in both unstructured 250 nm BSTS film (left column) and nanostructured BSTS metamaterial (right column) ) at 3 different angle of incidence, $\theta= 45°$ (top row), $\theta=0°$ (centre row) and $\theta=-45°$ (bottom row), with artificially increased chirality of the optical permittivity, $\varepsilon_r$, of the BSTS; the distinct behavior observed in the unstructured BSTS and the BSTS metamaterial matches remarkably well with the measured photocurrent, indicating how the chirality of the surface carriers and the increased absorption given by the metamaterial result in a giant increase of the CPGE; the $|E|^2$ is integrated in a 3 nm thin slab at the surface of the film. To better visualize the polarization ($\varphi$) dependence of both experimental data and numerical model, we subtract the polarization independent background from each curve and normalize them.

The electromagnetic modelling (Fig. 4b) is in excellent agreement with the experimental HDPC result (Fig. 4a), corroborating the link between anisotropic optical absorption at the BSTS surface and selective spin-photocurrent generation. The normalized polar plots in Fig 4a and 4b provide an immediate visualization of the prominence of linear components (4 lobes) versus circular components (2 lobes) in unstructured and nanostructured BSTS. While at normal incidence the BSTS metamaterial does not produce any notable effect, at oblique incidence it filters out linear components, distilling the response to circularly polarized light illumination, and leading to a giant enhancement of the chiral photocurrent. The degree of chirality predicted by the simulations, according to the same Eq. 2, is $\rho_{circ}$=0.18 for the unstructured BSTS, and $\rho_{circ}$=0.89 for the BSTS metamaterial, in excellent agreement with the values obtained experimentally (refer also to Supplementary Section IV, **Fig. S6**). Furthermore, in both experiments and simulations, illumination from mirror directions of incidence (with respect to the normal) yields opposite phases, just as expected for extrinsic chirality.

To conclude, the hitherto unrecognized ability of metamaterials to control surface transport in topological insulator by structural design provides a powerful toolbox to bridge the gap between nanophotonics and spin-electronics. We have shown that resonant nanostructures can be used to induce giant enhancement of the extrinsic chiral photocurrent response of a topological insulator. We argue that this approach could be used to exert control over spin-transport properties of other classes of quantum and topological materials (e.g. Weyl semimetals, van der Waals heterostructures), and find application in integrated spin-polarized photodetectors that are in great demand for ultrasensitive chiral molecular sensing and quantum opto-spintronic devices, where polarization and entanglement could be transferred from photons to electron spins.

## Methods

**Device fabrication**: $Bi_{1.5}Sb_{0.5}Te_{1.8}Se_{1.2}$ single crystals were grown using a modified Bridgeman method. This particular BSTS stoichiometry yields large surface to volume conductivity, so that transport in nanometric thin flakes is surface-dominated[22]. BSTS flakes were mechanically exfoliated and transferred from the bulk crystals onto a $SiO_2$ (285 nm)/*P*-Si substrates. Electrical contacts (Cr/Au 5/50 nm) for photocurrent measurements were created on the devices by electron-beam lithography (EBL) and thermal evaporation. Square rings metamaterials were carved on the BSTS flake between the contacts by focused ion-beam (FIB) milling. All fabrication steps were performed minimizing the exposure of BSTS to the electron and ion beams. SEM images were acquired after photocurrent measurements.

**Helicity dependent photocurrent measurement**: Photocurrent measurements were performed at room temperature, illuminating the devices with continuous wave laser ($\lambda$=532 nm) and with no applied bias. The linearly polarized laser beam was modulated at frequency of 137 Hz by an optical chopper before passing through a $\lambda/4$ retarder, and focused to a spot size of ~200 μm diameter at the centre of the electrodes. The experimental setup allowed to continuously vary the polarization of incident light from linear (s-polarized) to circular (RCP and LCP) by rotating the $\lambda/4$ wave plate. The incident light polarization was calibrated by a polarimeter. The photocurrent was measured with a lock-in amplifier referenced to the light modulation frequency. In all samples, the photocurrent was found to be linearly dependent on excitation intensity (refer to Supplementary Information, **Fig S1**). All the measurements were performed at a constant illumination intensity of 10 Wcm$^{-2}$.

**Electromagnetic simulations:** 3D electromagnetic simulations were performed using COMSOL Multiphysics, using the experimental relative permittivity of BSTS, $\varepsilon_r$, obtained by ellipsometry. The anisotropic response along the *x* direction was introduced by assigning non-zero values to the off-diagonal terms of the permittivity tensor, $\varepsilon_{r(yz,zy)}= \pm j*\varepsilon_{r(xx,yy,zz)}$. The incident wave polarization rotation, reproducing the experimental arrangement of polarizer and QWP, was obtained by introducing a φ dependence to both, *s* and *p* components of the electric field with a π/2 phase retardation between the two. The electric field intensity plotted in the graphs was obtained by integrating the $|E|^2$ within a 3 nm thin slab from the surface.

## Acknowledgements

The authors would like to acknowledge Justin Song for insightful discussions on fundamental processes underlying HDPC in Dirac materials, Alexander Dubrovkin, Guanghui Yuan and Syed Aljunid for technical consultations, and Wang Lan for providing the BSTS crystal. This research was supported by Singapore Ministry of Education (Grant No. MOE2016-T3-1-006 (S)) and the Singapore National Research Foundation, Prime Minister's Office, under its Quantum Engineering Programme (Grant No.: QEP-P1). M.E. acknowledges the 100 Foreign Talents Project in Jiangsu Province (JSA2016003) and the National Natural Science Foundation of China (NSFC 1774170) for travel support.

## Author contributions

C.S., M.E. and G.A. conceived the original idea. X.S. developed experimental setup with initial assistance from M.E. and performed all HDPC measurements. G.A. and X.S. developed the device fabrication process (G.A. fabricated the metamaterials and X.S. fabricated the devices). G.A. performed the electromagnetic simulations. H.N.S.K. did ellipsometric measurements and extracted the optical constants. X.S., G.A. and C.S analysed the data and drafted the manuscript. All authors contributed to discussion and revision of the manuscript. C.S. and N.Z. supervised the work.

## Competing interests

The authors declare no competing interests.

## Additional Information

**Supplementary Information** is available for this paper.

**Correspondence and requests for materials** should be addressed to C.S.